\documentclass[12pt,preprint]{aastex}
\usepackage{emulateapj5}
\def\msolar{\ifmmode{M_\odot}
    \else{$M_\odot$}\fi}
\def\lsolar{\ifmmode{L_\odot}
    \else{$L_\odot$}\fi}
\def\msun{\ifmmode{M_\odot}
    \else{$M_\odot$}\fi}
\def\lsun{\ifmmode{L_\odot}
    \else{$L_\odot$}\fi}
\newcommand\ra[4]{\ifmmode{#1^h#2^m#3\fs #4}
    \else{$#1^h#2^m#3\fs #4$}\fi}
\newcommand\dec[4]{\ifmmode{#1\arcdeg #2\arcmin #3\farcs #4}
    \else{$#1\arcdeg #2\arcmin #3\farcs #4$}\fi}
\newcommand\nra[4]{\ifmmode{#1\ #2\ #3.#4}
    \else{$#1\ #2\ #3.#4$}\fi}
\newcommand\ndec[4]{\ifmmode{#1\ #2\ #3.#4}
    \else{$#1\ #2\ #3.#4$}\fi}

\newcommand\scnot[2]{\ifmmode{#1\times 10^{#2}}
    \else{$#1\times 10^{#2}$}\fi}
\newcommand\jeq[2]{\ifmmode{{J=#1\rightarrow #2}}
    \else{{$J=#1\rightarrow #2$}}\fi}
\newcommand\feq[2]{\ifmmode{{F=#1\rightarrow #2}}
    \else{{$F=#1\rightarrow #2$}}\fi}
\newcommand\cm[1]{\ifmmode{~{\rm cm}^{#1}}
    \else{~${\rm cm}^{#1}$}\fi}
\newcommand\jkeq[6]{\ifmmode{{J_{K_{-1}K_1}=#1_{#2#3}\rightarrow
	#4_{#5#6}}}	
     \else{{$J_{K_{-1}K_1}=#1_{#2#3}\rightarrow #4_{#5#6}$}}\fi}
\newcommand\frjk[6]{\ifmmode{{\frac{I(#1_{#2#3})}{I(#4_{#5#6})}}}
     \else{{$\frac{I(#1_{#2#3})}{I(#4_{#5#6})}$}}\fi}
\def\kms{${\rm km\ s}^{-1}$}
\def\gsim{{\scriptscriptstyle \stackrel{>}{\sim}}}
\def\lsim{{\scriptscriptstyle \stackrel{<}{\sim}}}
\newcommand\intens{\ifmmode{\int T_{\rm R}^*dv}
	\else{$\int T_{\rm R}^*dv$}\fi}
\newcommand\anttemp{\ifmmode{\int T_{\rm A}dv}
	\else{$\int T_{\rm A}dv$}\fi}

\def\xe{$X_{e^{-}}$}

\def\dtoh{${\rm [D]}/{\rm [H]}$}

\def\ratio{${\rm NH_2D/NH_3}$}

\def\rdcop{{${\rm DCO^+}/{\rm HCO^+}$}}
\def\rhcop{{${\rm DCO^+}/{\rm HCO^+}$}}
\def\rhcn{${\rm DCN}/{\rm HCN}$}

\def\nhthr{{${\rm NH}_3$}}
\def\ammonia{{${\rm NH}_3$}}
\def\nh2d{{${\rm NH}_2{\rm D}$}}

\def\htwco{${\rm H}_2{\rm CO}$}
\def\formal{{${\rm H}_2{\rm CO}$}}

\def\water{{${\rm H_2O}$}}
\def\htwo{${\rm H_2}$}
\def\hydrogen{${\rm H_2}$}

\def\c18o{${\rm C^{18}O}$}

\def\htwo18o{${\rm H_2^{18}O}$}

\def\hthrp{${\rm H_3}^+$}
\def\htwdp{${\rm H_2D}^+$}

\def\ch2dp{${\rm CH}_2{\rm D}^+$}
\def\ntwohp{${\rm N_2H}^+$}
\def\n2dp{${\rm N_2D}^+$}

\def\hcop{{${\rm HCO}^+$}}
\def\dcop{{${\rm DCO}^+$}}

\begin{document}
\title{Deuterated Ammonia in Galactic Protostellar Cores}

\author{Ronak Y. Shah\altaffilmark{1}}
\email{Email: shah4@astro.uiuc.edu}
\and
\author{Alwyn Wootten}
\email{Email: awootten@nrao.edu}
\affil{National Radio Astronomy Observatory, Charlottesville, VA
22903}
\altaffiltext{1}{Current Address:  University of Illinois at
Urbana-Champaign, 1002 W. Green Street, Urbana, IL  61801}
\received{2000 December 15}
\revised{2001 February 16}
\accepted{2001 February 21}
\cpright{PD}{2001}

%%
%% Abstract
%%
\begin{abstract}
We report on a survey of \nh2d\ towards protostellar cores in
low--mass star formation and quiescent regions in the Galaxy.
Twenty--three out of thirty-two observed sources have significant
($\gsim 5\sigma$) \nh2d\ emission.  Ion--molecule chemistry, which
preferentially enhances deuterium in molecules above its cosmological
value of \scnot{1.6}{-5} sufficiently explains these abundances.
\ratio\ ratios towards Class 0 sources yields information about the
``fossil remnants'' from the era prior to the onset of core collapse
and star formation.  We compare our observations with predictions of
gas--phase chemical networks.
\end{abstract}


%%
%% Introduction
%%
\section{Introduction}

Deuterium in molecules is enhanced above the cosmological
\dtoh\ value of $1.6^{+0.14}_{-0.19}\times 10^{-5}$ \citep{linsky1995}
by several orders of magnitude in the cold environs of the
interstellar medium (ISM).  Two theoretical pathways can account for
the abundance of these deuterated molecules.  Deuterium binds more
strongly in molecules than do less massive hydrogen atoms, especially
at low temperatures, in gas--phase reactions between parent species
and e.g. ${\rm H_3^+}$ and \htwdp\ \citep{dl1984,watson1974} and other
similar molecular ions in dense ($n>10^4{\rm cm^{-3}}$) gas.  The
qualitative understanding that reactions specifically with ${\rm
H_3^+}$ and \htwdp\ have been enhanced by recent detections
\citep{geballeoka1996,stark1999}.  Their abundances are consistent
with requirements of gas--phase ion--molecule chemistry predictions of
most species, including deuterium isotopes.  Reactions on grains will
also contribute to deuterium enhancements.  Deuteration has a smaller
activation energy than does hydrogenation on grain surfaces.
Additionally the more massive deuterium will desorb less quickly from
grain surfaces than hydrogen, increasing the interaction time for
potential grain--induced deuteration \citep{tielens1983}.  Such
predictions are considered important for the high fractionation ($\sim
10^{-2}$) observed for ${\rm HDCO/H_2CO}$ \citep{loren1985}, ${\rm
NH_2D/NH_3}$ \citep{walmsley1987}, and ${\rm CH_3OD/CH_3OH}$
\citep{mauersberger1988} towards the hot core Orion-KL.  In such
regions, high temperatures quench the ion--molecule deuterium
fractionation reactions by rapid destruction of ${\rm H_2D^+}$ and
faster reactions (lower energy barriers) for more abundant species.
High observed deuterium fractions towards these sources provide a
chemical ``fingerprint'' of the physical conditions in the
pre--protostellar gas.  Comparisons of deuterated molecules to their
more abundant isotopes may, then, provide meaningful insight into the
evolution of the ISM.

Deuterated ammonia, \nh2d, is a useful probe of protostellar sources
because (1) it can be easily compared to \ammonia, a well--used tracer
of temperature and dense gas condesations in the ISM; (2) it possesses
easily observable millimeter transitions; (3) it has hyperfine
components that can be easily resolved and used to estimate column
density; and (4) its abundance can be compared with chemical models
which predict variations with temperature, density and evolutionary
state.  \nh2d\ was first observed by \citet{turner1978} in Sgr B2 and
\citet{rodkuip1978} in Orion KL.  These observations of hot core
regions yielded ammonia deuterium fractionation levels, \ratio, of
order a few times $10^{-3}$, significantly higher than the local ISM
\dtoh\ value of \scnot{1.6}{-5}.  \citet{walmsley1987}
suggested that longer time-scales for deuterium interaction on grain
surfaces (versus that of lighter hydrogen atoms) can generate larger
deuterated molecular abundances.  The high temperatures in these cores
sublimate or photo-evaporate grain mantles, providing the high
observed \ratio\ levels.  \citet{olberg85} investigated this
phenomenon in the colder, less evolved sources L183 and towards the
\ammonia\ column density peak in TMC1, TMC1~\ammonia, where low
temperatures minimize any surface contributions and gas--phase
chemistry dominates molecule formation.  They found \ratio
(L183)=0.05, but only an upper limit towards TMC1~\ammonia.  Tin\'{e}
et al. (2000) revisit the difference in ammonia fractionation for L183
and TMC1.  They find \nh2d\ detections towards both sources, with L183
(${\rm NH_2D/NH_3}\sim 0.1$) having 5 times the ammonia fractionation
level of TMC1 (${\rm NH_2D/NH_3}=0.02$).  Their gas--phase network
based on the reaction rates of Millar et al. (1991) sufficiently
explains the abundances and fractionation of ammonia.  Saito et
al. (2000) invoke grain processes to differentiate star--filled and
star--less ammonia cores with and without \nh2d\ detections,
respectively.  They made this conclusion despite the fact that Saito
et al. (2000) measured the strong fractionation towards L183, with a
value similar to ours and Tine et al.  L183 is a well known starless
core which lacks any source of radiation to sublimate grains.  Some of
the discrepancy in the interpretation of the origin of ${\rm NH_2D}$
may result from the source samples.  The listed studies have
concentrated on either too few sources or sources over a small range
in physical conditions.

We present in this paper an experiment to address the deuterium
enhancements of ammonia, ${\rm NH_2D/NH_3}$, towards dark cores and
Class 0 sources.  The former are often termed ``pre--protosteller''
since most do not possess strong continuum detectable by instruments
such as {\it IRAS}.  In such environs, ion--molecule chemistry
fractionates neutrals and ions such as ammonia and \hcop\
\citep{wootten1987} and dominates surface deuteratated components
desorbed back into the ISM.  Most of the grain ice mantles remain
frozen since the ambient temperatures are significantly less than
90~K, the sublimation temperature of the major ice constituent,
\water.  Class 0 sources are cold dense cores with spectral energy
distributions (SEDs) peaked in the sub-millimeter (${\rm T_D}\sim
20-50$ K) and highly collimated outflows
\citep{awb93}.  \citet{blake1995} showed that Class 0 sources such
as IRAS4 contain a large number of hydrogenated species and
long--chain molecules that can only be produced by grain catalyzation.
Thus, Class 0 sources represent an ideal location for separating
gas--phase effects from grain chemistry.  The \ratio\ ratio of these
sources will reflect what, if any, additional deuteration has occured
on the grains.
%
%However, note recent detections of ${\rm D_2CO}$
%towards IRAS~16293, \citet{castets1999} and ${\rm NHD_2}$ towards L183
%\citep{roueff2000} which point to grain species even in sources much
%colder than Orion where they were originally detected
%\citet{turner1978}.  

Our observations of \nh2d\ also reveal that four (and possibly six)
sources possess asymmetric line profiles.  One does not expect to see
such complex line profiles from relatively low abundance (and,
therefore, low optical depth) species such as \nh2d.  Line--of--sight
self--absorption or kinematic effects (e.g. multiple cores within the
telescope beam) can account for such a line profile.  In this work, we
assert that gas--phase chemistry adequately accounts for observed
deuterium enhancements in ammonia.  We present a radiative transfer
analysis using (multi-layer) microturbulent models in $\S$3, and
discuss how ion--molecule reactions can, in fact, effectively explain
the observations, including self--absorbed profiles in section $\S$5.
We discuss our main conclusions in section $\S$6.

%%
%% Observations
%%
\section{Observations \label{obs}}

\subsubsection{The \nh2d\ molecule}

\nh2d\ is a slightly asymmetric top molecule whose rotational levels
are split by inversion doubling, as in its more well studied symmetric
isotopomer, ${\rm NH}_3$, but the asymmetry mixes the rotation and
inversion. The primary dipole moment, $\vec{\mu_c}=1.4652$~D,
induces rotation--inversion transitions which are in turn split by the
$^{14}$N nuclear quadrupole moment into five hyperfine components
\citep{bester1983}.  The inversion--split $1_{01}$ and $1_{11}$
levels, corresponding to the $\mu_c$ dipole moment, produce a
symmetric and anti--symmetric transition, one at 85.926263 GHz
(ortho--\nh2d; $v=0\rightarrow 1$) and the other at 110.153599 GHz
(para--\nh2d; $v=1\rightarrow 0$) (see Table \ref{nh2dfreqtable}).
The relative nuclear statistical weights taken from \citet{bester1983}
favor the para--\nh2d\ transition by a factor of 3.  As in ammonia,
one can compute the optical depths and column densities using ratios
of the hyperfine transitions \citep{mwm92,ho83}.  A weaker dipole
moment, $\vec{\mu_a}=-0.1848$~D, produces pure rotational transitions
within a given inversion level.  It is responsible for the ground
state transitions, $\jeq{1_{01}}{0_{00}}$ levels at 332.82251~GHz
($v=0$) and 332.78189~GHz ($v=1$).
\vspace{2cm}

\subsection{NRAO 12--meter Observations}

\begin{figure}
\epsscale{0.7}
\plotone{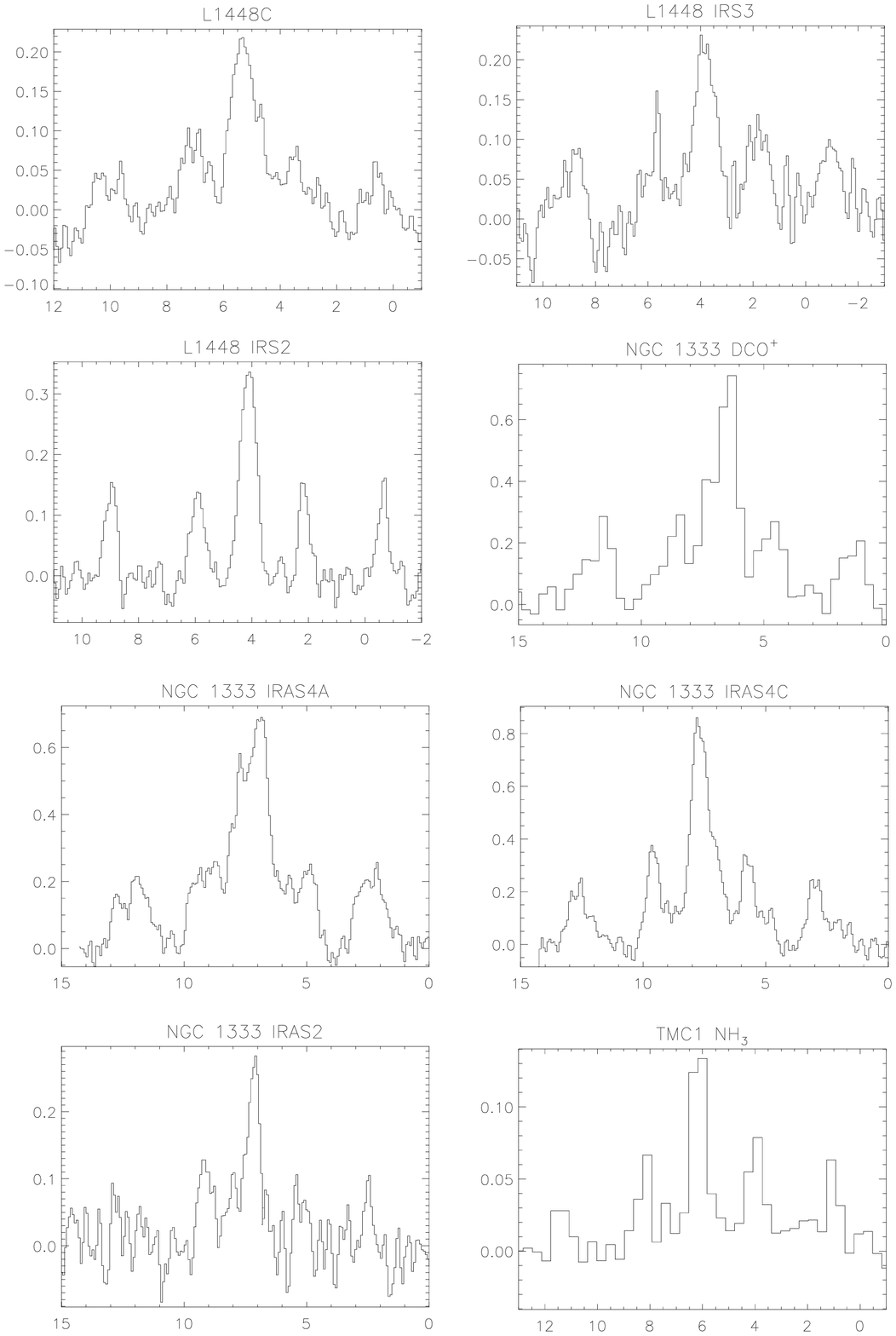}
\caption{85.9 GHz \nh2d\ spectra.  The spectral resolution
is 0.13 \kms/channel except for NGC1333 \dcop\ and TMC1~\ammonia where
it is 0.35 km/s.  A linear baseline is removed from the data,
revealing the hyperfine components.  \label{fig:plot86_1}}
\end{figure}

\begin{figure}
\plotone{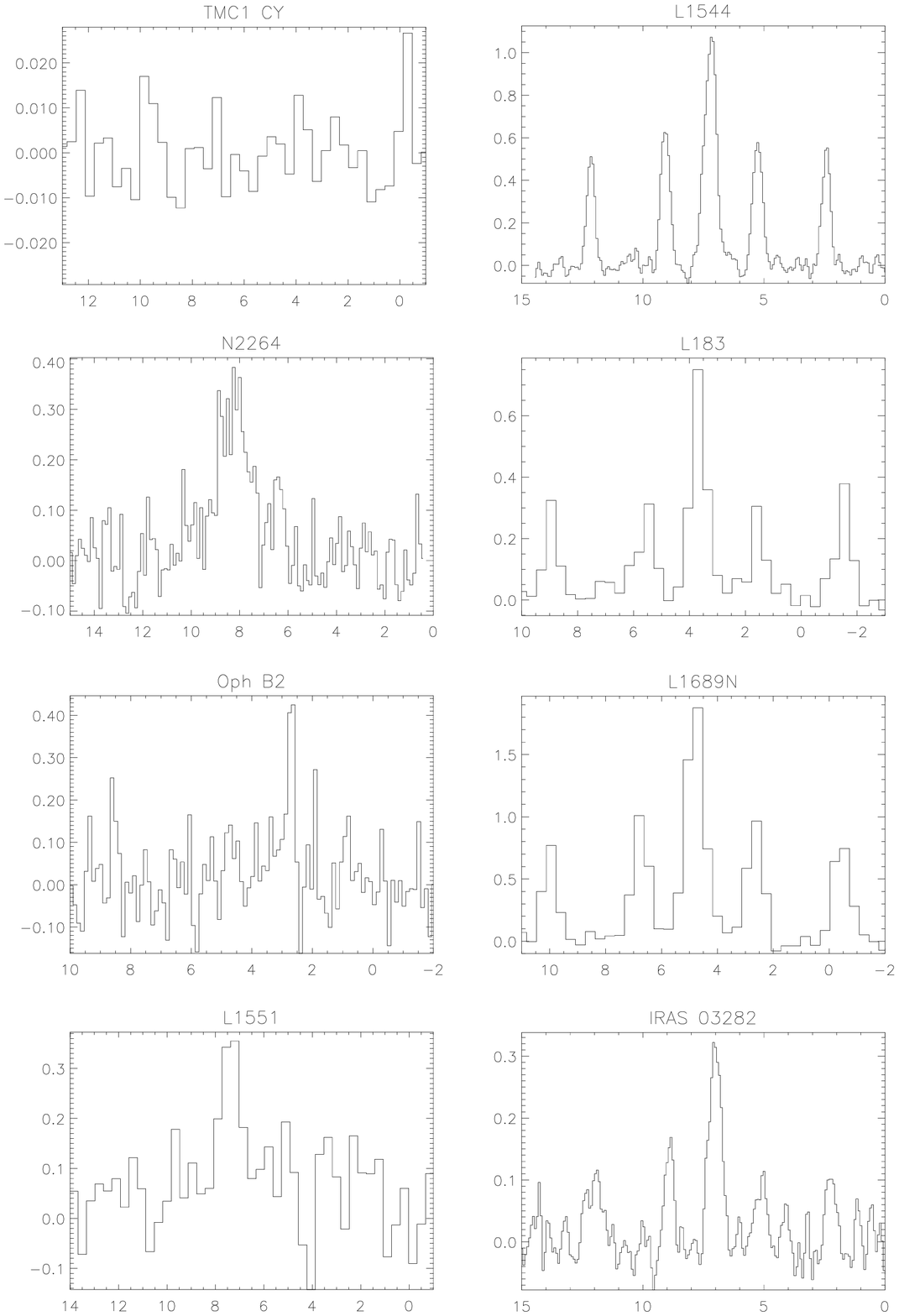}
\caption{Same as figure \ref{fig:plot86_1}.  Here TMC1~CY, L183,
L1689N and L1551 are observed with the coarser 0.35 km/s
resolution. \label{fig:plot86_2}}
\end{figure}

\begin{figure}
\plotone{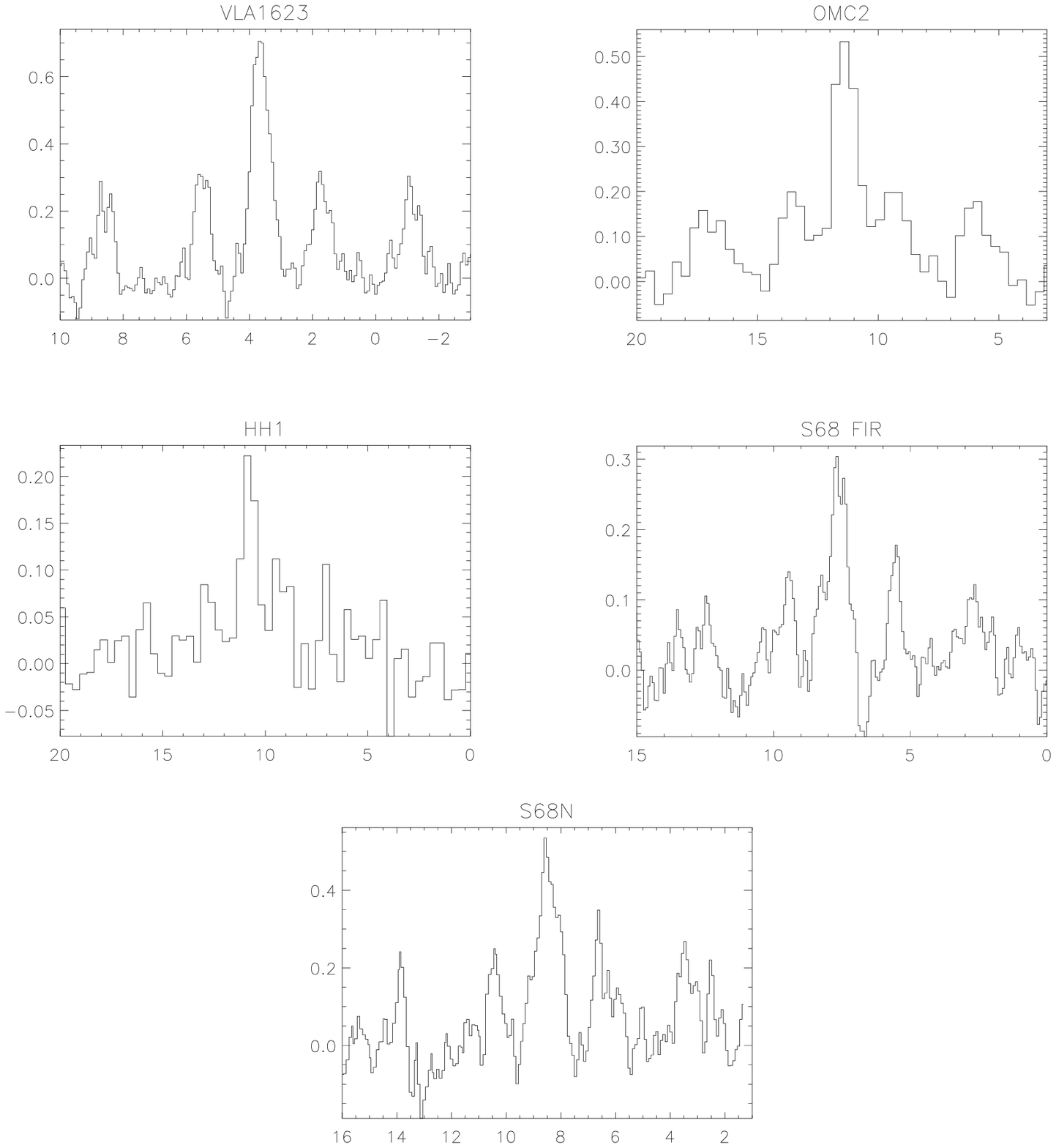}
\caption{Same as figure \ref{fig:plot86_1}.  Here TMC1~CY, L183,
L1689N and L1551 are observed with the coarser 0.35 km/s
resolution. \label{fig:plot86_3}} 
\end{figure}

We observed the 85.93GHz $J_{K_{-1}K_1}=1_{11}^a\rightarrow 1_{01}^s$ and
110.15~GHz $J_{K_{-1}K_1}=1_{11}^s\rightarrow 1_{01}^a$ rotation--inversion
transitions of \nh2d\ using the NRAO 12~meter\footnote{The National
Radio Astronomy Observatory is operated by Associated Universities,
Inc., under cooperative agreement with the National Science
Foundation.} telescope at Kitt Peak, Arizona (Figures
\ref{fig:plot86_1}, \ref{fig:plot86_2}, and \ref{fig:plot86_3}).  The
data are from several epochs, mostly dating from 1987
September, but a fair portion from 1997 September, using both the
filter banks and the hybrid digital spectrometer.  The recent data set
was observed with a two--channel SIS junction receiver tuned to
receive signals in a single side--band.  Each channel measured
orthogonal linear polarizations.  We utilized two banks of filters,
100 kHz and 250 kHz, each consisting of 256 channels, resulting in
0.27~\kms and 0.68~\kms per channel resolution.  We used the hybrid
spectrometer in the two--IF mode with 12.5 MHz bandwidth and 512
channels per IF, resulting in a spectral resolution of 24.4 kHz or
0.131\kms; both linear polarizations were averaged.  All data were
obtained in frequency switched mode, at a switching interval of 2.5
MHz.  The excellent quality of the high resolution data from the
hybrid spectrometer motivated us to use it exclusively for our
analysis presented here.  Filter bank data were generally used to
diagnose any systematic problems.

Sky dip measurements indicated zenith opacities of 0.01 to 0.4 when few
clouds were apparent.  The system temperatures varied between 180 and
300 K for all observations reported here.  Calibration was achieved by
the synchronous detection of an ambient temperature absorber and the
sky.  The observed line intensities were corrected for forward
scattering and spillover efficiency $\eta_{fss}=0.68$ to place the
data on a T$_R^*$ scale.  Column density calculations require an
additional main beam brightness correction of $\eta_m^*=0.95$.  The
beamsize of the 12m at 86 GHz is 90\arcsec, and 70\arcsec\ at 110 GHz.
Regular pointing checks performed with planets indicated a positional
accuracy better than 5\arcsec.

\subsection{CSO Observations of \nh2d}

We also report on the detection of the fundamental transition of
\nh2d, $J=1_{01}\rightarrow 0_{00}$, at 332.782~GHz (Ortho) and
332.822~GHz (Para), towards the dense core L1689N with the Caltech
Submillimeter Observatory (CSO).  The data were observed in orthogonal
linear polarizations.  The 1024--channel 50MHz acousto--optical
spectrometers were used.  The data were position switched with
30\arcmin\ east or west offsets.  The beamwidth at 333 GHz is
$20\arcsec$ and the main beam efficiency, determined from planet
observations, is 67\% at 345~GHz.

\subsection{NRAO 140--foot Observations}

\begin{figure}
\epsscale{0.9}
\plotone{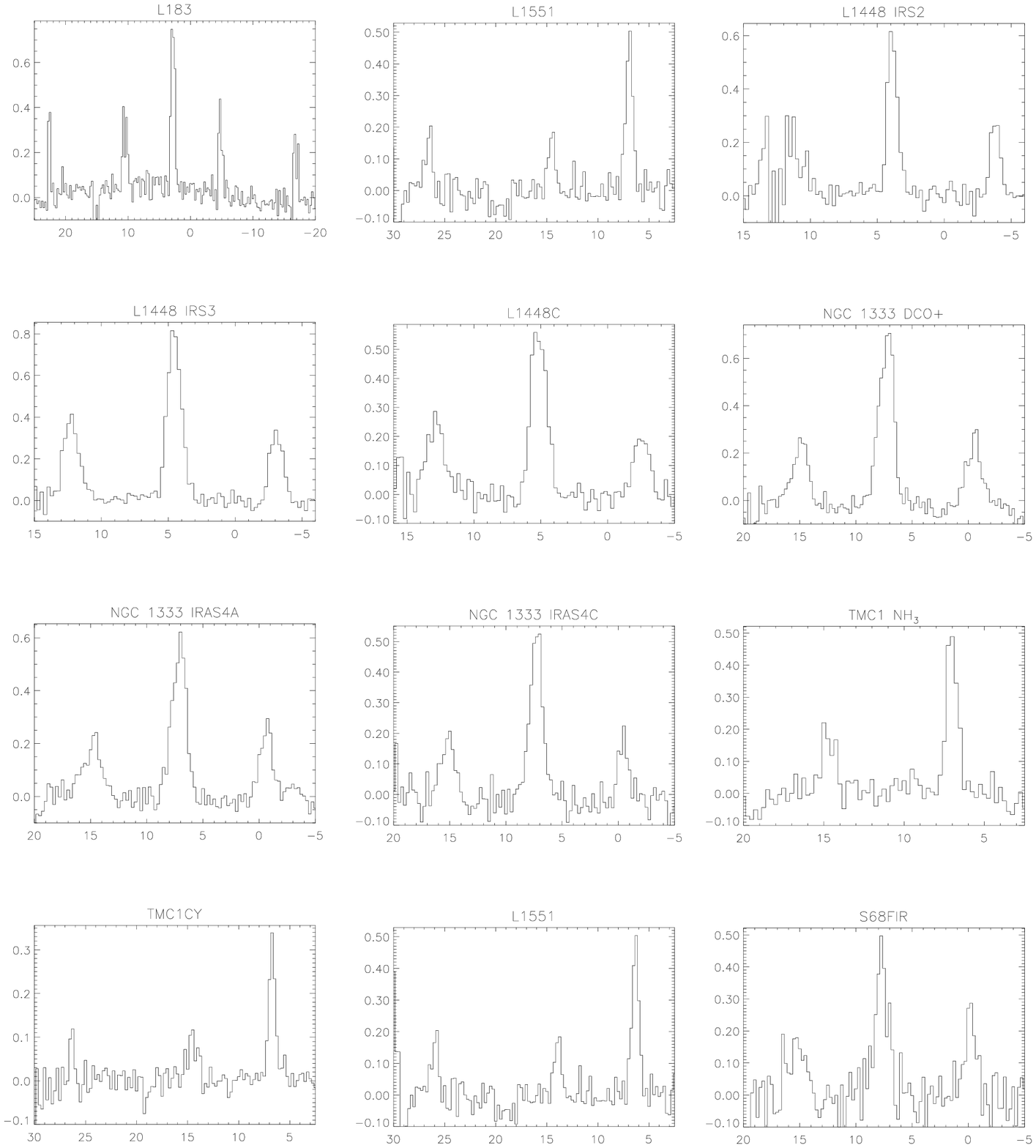}
\caption{23.6 GHz \ammonia\ spectra.  The spectral resolution
is 0.25 \kms/channel.  A linear baseline is removed from the data,
revealing the hyperfine components.  \label{fig:plotnh3_1}}
\end{figure}

Observations of the ammonia transitions were made during 1998 June at
the 43m telescope of the NRAO in Green Bank, West Virginia (Figure
\ref{fig:plotnh3_1}).  The $(J,K)=(1,1)$
(23.694506~GHz) and $(J,K)=(2,2)$ (23.722634~GHz) transitions were
observed to sensitive limits in all of our sources.  All of the data
were observed in orthogonal linear polarizations.  The 1024 channel
Mark IV autocorrelator gave a spectral resolution of 0.25
\kms\ per channel.  The data were frequency switched at an interval of
2.5 MHz.  The variation of gain with elevation was reduced by use of
the lateral focusing stage.  The beamwidth at 23.7 GHz is $\sim
1\arcmin$ and the beam efficiency is 30\%.  This beam width is similar
to that of the 12--meter observations.  Thus, differential beam
dilution is not considered when comparing the \nh2d\ and \nhthr\ column
densities, assuming that both isotopes emit from similar volumes of
gas.  Line intensities are measured in units of T$_R$*.  The forward
scattering and spillover efficiency, $\eta_{\rm fss}$, is 0.63 (Loren,
Evans, and Knapp 1979).

%%
%% Results
%%

\section{Results \label{nh2dsurveyresults}}

\subsection{\ammonia\ and \nh2d \label{ammonia}}

We determined column densities for both \nhthr\ and \nh2d\ using a
microturbulent radiative transfer model.  We utilize \formal\ as a
measure of source physical conditions.  This explictly assumes that
\formal\ and ammonia overlap.  In general, \formal\ is excited in more
dense regions than \ammonia.  Thus, temperature estimates from both
species may not be cospatial in a given source.  With this caveat
in mind, we discuss below how physical conditions were determined.

\subsubsection{Estimates of Source Physical Conditions\label{formaldehyde}}

Chemical models predict significant variations of molecular abundances
with density, temperature, and evolutionary state of the cloud.  We
model the temperature and density structure of each source using
\ammonia\ observations and formaldehyde observations from
\citet{wootten2001}.  For clarity we discuss some issues involved with
\formal\ radiative transfer modeling below.  \formal, an asymmetric
rotor molecule, is an important probe of molecular cloud physical
conditions because of:

\begin{itemize}
\item a large number of transitions closely spaced in frequency
but well separated in energy, possessing a coupled sensitivity to
kinetic temperature {\it and} volume density when measured through
appropriately chosen line ratios.
\item a large, uniform, and ubiquitous abundance in the ISM
\citep{mangum1990}. 
\item multiple transitions through only a few receiving systems,
minimizing calibration uncertainties.
\end{itemize}

\noindent One must compare the various transitions to determine if an
isothermal, iso--density cloud model or one with more elaborate
physical conditions is necessary to reproduce the observed \formal\
transitions.  

Several steps are needed to determine basic physical parameters for
the sources.  The ratio of the \ammonia\ (2,2) and (1,1) measurements
provides lower limits to the temperature of the cold, extended
envelopes of each source.  We use published data with similar
resolution to supplement some of our (2,2) data with poor
signal--to--noise (see column 7 of Table \ref{dammntottab}).  Next, we
use the formaldehyde data which provides sensitivity to the density
and the presence of any warm gas.  A comparison of many \formal\ line
ratios using a Large Velocity Gradient model provides initial
estimates for both density and temperature \citep{mangum1993}.  Using
these initial estimates, we then employ a more robust microturbulent
radiative transfer model (one in which the size scale of velocity
turbulence is small when compared to the photon mean--free path) to
iteratively calculate the line intensities of the lower opacity
para--formaldehyde species and obtain the best fit temperature and
density model which reproduces the observed brightness temperatures.
In some regions, multiple layers must contribute to the observed
spectrum.

In general, low excitation transitions of \formal\ (e.g.
$J_{K_{-1}K_1}=1_{01}\rightarrow 1_{00}$ are compared to estimate the
temperature and density in the cold region.  Alternatively, a
relatively warmer region tends to produce emission from higher energy
transitions (e.g. $J_{K_{-1}K_1}=5_{05}\rightarrow 4_{04}$).  For
example, towards IRAS4A, \citet{wootten2001} find both low excitation
($\int T_R^*dV [J_{K_{-1}K_1}=1_{01}\rightarrow 1_{00}] = 4.21\pm
0.09$~K~\kms) and high excitation ($\int T_R^*dV
[J_{K_{-1}K_1}=5_{05}\rightarrow 4_{04}] = 3.63\pm 0.15$~K~\kms) lines
of \formal, indicating that a layer of high density, warm gas coexists
with colder, low density gas.

We use these values for n and T as input parameters into a
microturbulent model for \nh2d\ and \ammonia.  The partition function
is determined over all energy levels in order to eliminate errors due
to high temperature approximations.  With the size of the region set
by our beam, we essentially have one free parameter, the column
density, to predict the integrated intensities.  We match the
intensities and line profiles of all of the hyperfine components of
\nh2d\ and \ammonia\, along with estimates of the optical depth from
the following equation:

\begin{eqnarray}
\frac{\rm T_B(m)}{\rm T_B(s)} & = & 
\frac{1-exp[-\tau(m)]}{1-exp[-a\tau(m)]} \label{ammtaueq}
\end{eqnarray}

\noindent to determine the best fit column density.  In equation
\ref{ammtaueq}  $m$ and $s$ stand for main and satellite hyperfine lines,
respectively; ${\rm T_B}$ is the observed brightness temperature;
$\tau$ is the optical depth; and $a$ is the ratio of the satellite to
main hyperfine intensities.  We list all results in Tables
\ref{dammntottab} and \ref{ammntottab} for the deutero--ammonia and
ammonia column densities.  $5\sigma$ upper limits for the line
intensities and column densities are listed in Table \ref{upperlims}.
In Tables \ref{ammfracsingle} \& \ref{multilayer} we list the ammonia
fractionation for single--temperature and multiple--layer models,
respectively. 

The quality of our \ammonia\ data was not sufficient for reasonable
comparison with our \nh2d\ in all cases.  In general, not all of the
hyperfine components were obtained in one spectrum because of hardware
errors in the velocity offsets.  This required us to carefully compare
all column density estimates based on only a subset of the ammonia
hyperfine components, with data of similar spatial and spectral
resolution found in the literature.  We list all references for
ammonia data in Table \ref{ammntottab}.  In all cases we find good
agreement between our own \ammonia\ column density estimates and those
available in the literature.  This lends confidence to our estimates
of ammonia deuterium fractionation.

\subsubsection{Para--to--Ortho Ratio for \nh2d.}

In Table \ref{orthopara} we list the Ortho--to--Para ratio for the 86
and 110 GHz transitions of \nh2d\ for eight sources.  If one compares
the statistical weights, the 110~GHz line strength is three times
weaker than the 86~GHz transition.  Different excitation conditions
may vary this ratio slightly.  Direct comparison of the integrated
line strengths indicated that the ratio
$\intens[86~GHz]/\intens[110~GHz]$ is consistent with 3 for all
sources we observed in both lines.  We also determine the column
density of ortho--\nh2d for comparison to para--\nh2d.  Again, we find
that the ratios of the column densities in the upper levels of both
transitions are consistent with three.

\subsection{Self--absorption \label{selfabssec}}

We find asymmetric spectral features indicative of an inwardly
increasing temperature gradient in low--mass protostars in the 86~GHz
NH2D transitions towards NGC~1333 IRAS4A, L1448C, S68~FIRS1, and
possibly in NGC~1333 IRAS4C and S68N.  Very high resolution profiles
show similar spectral signatures in \htwco, \ntwohp, \hcop, and CS
(Mardones et al. 1997; Gregersen et al. 1997).  Line asymmetries or
skewness were computed using the following dimensionless quantity:

\begin{eqnarray}
Skewness & = & {\rm \Sigma T(V-V_{LSR})^3\Delta V/(\Sigma T\Delta V)}
\times \\ 
         & = & {\rm \left\{(\Sigma T(V-V_{LSR})^2\Delta
V)/(\Sigma T\Delta V)\right\}^{-3/2}} \nonumber
\end{eqnarray}

\noindent Here, we take a weighted sum over each of the hyperfine
components to obtain an average skew of the entire \nh2d\ line
profile.  For a line that possesses most of its emission at velocities
less than the source's systemic velocity (``blue-ward'' asymmetric),
the skewness is negative; it is positive for a similar ``red-ward''
asymmetry.  We list our calculations for those sources with
significantly asymmetric line profiles in Table \ref{tinfall}.  This
method indicates that IRAS4A and L1448C possess the largest blue
asymmetry.  The \nh2d\ spectra of IRAS4C and S68N show strong
blue-ward behavior as well.  The relative uncertainty in the S68~FIRS1
data makes quantifying an asymmetry difficult.  No sources show
significant ($\gsim 1\sigma$) red-ward asymmetries.

%%
%% Chemistry
%%
\section{Deuterium Chemistry and Ammonia\label{chemistry}}

Gas-phase synthesis of \nh2d\ is strongly supported by observational
evidence for \hthrp\ \citep{mccall1998,geballeoka1996} and \htwdp\
\citep{stark1999} ion--molecule chemistry model predictions, and
limits to desorption mechanisms in the molecular gas towards our
observed sources.  We discuss grain desorption mechanisms in
$\S$\ref{Discussion}.  For clarity, we briefly discuss the general
ion--molecule chemistry responsible for \nh2d\ and \ammonia\
production.  We use the detailed models of \citet{rm} and
\citet{turner2001} to compare our data to in $\S$\ref{Discussion}.  We
refer the reader to those sources for a more extensive discussion.
Using the most current set of reaction rates from the UMIST database
\citep{millar1997}, \citet{rm} determined the most important chains
for the deuterium injection into the ammonia production process.  The
three general schemes considered in the literature are:

\begin{eqnarray}
{\rm H_2} + {\rm N^+} & \rightarrow & {\rm NH}^+   + {\rm H} \label{eqone} \\
{\rm H_2^+} + {\rm N} & \rightarrow & {\rm NH}^+   + {\rm H} \label{eqtwo} \\
{\rm H_3^+} + {\rm N} & \rightarrow & {\rm NH_2}^+ + {\rm H} \label{eqthree}
\end{eqnarray}

\noindent Reaction \ref{eqone} is often rejected for \ammonia\
production because experiments performed at very low temperatures
indicate a reaction endothermicity of 85K \citep{marquette1985} for
thermal conditions.  The corresponding reaction with HD, on the other
had, has an endothermicity of 16K.  This reaction also has the problem
that ${\rm N^+}$ is not very abundant.  \citet{herbst1987} suggests
that a nonthermal formation mechanism, where ${\rm N^+}$ obtains
additional translational energy in the reaction

\begin{eqnarray}
{\rm N_2} + {\rm He^+} & \rightarrow & {\rm (N^+)^*} + {\rm N} + {\rm
He} \label{eqfour}
\end{eqnarray}

\noindent can provide the necessary abundance of ${\rm N^+}$ for
equation \ref{eqone}.  A standard set of laboratory-determined
reaction rates have not been determined.  The range of rates lead to a
factor of 10-100 variation in predicted ammonia abundances.  Reaction
\ref{eqtwo} works well only for the formation of \ammonia, since the
deuterium equivalent has an energy barrier and limited abundance of
${\rm HD^+}$.  Reaction \ref{eqthree} has received much recent debate
since its rate was remeasured by \citet{scott1997} to be sufficient
for ammonia (and, assuming the same rate, \nh2d) production.  However,
this contradicted earlier theoretical work which indicated that
double--proton transfer was too inefficient \citep{huntress1977}, even
if favored at cold temperatures.  A more recent measurement for the
rate of equation \ref{eqthree} indicates that the result of
\citet{scott1997} was erroneous.  The authors no longer consider it a
viable channel for ammonia production.

%%
%% Discussion
%%
\section{Discussion \label{Discussion}}

\subsection{Comparisons with gas--phase model predictions of \nh2d.}

\begin{figure}
\plotone{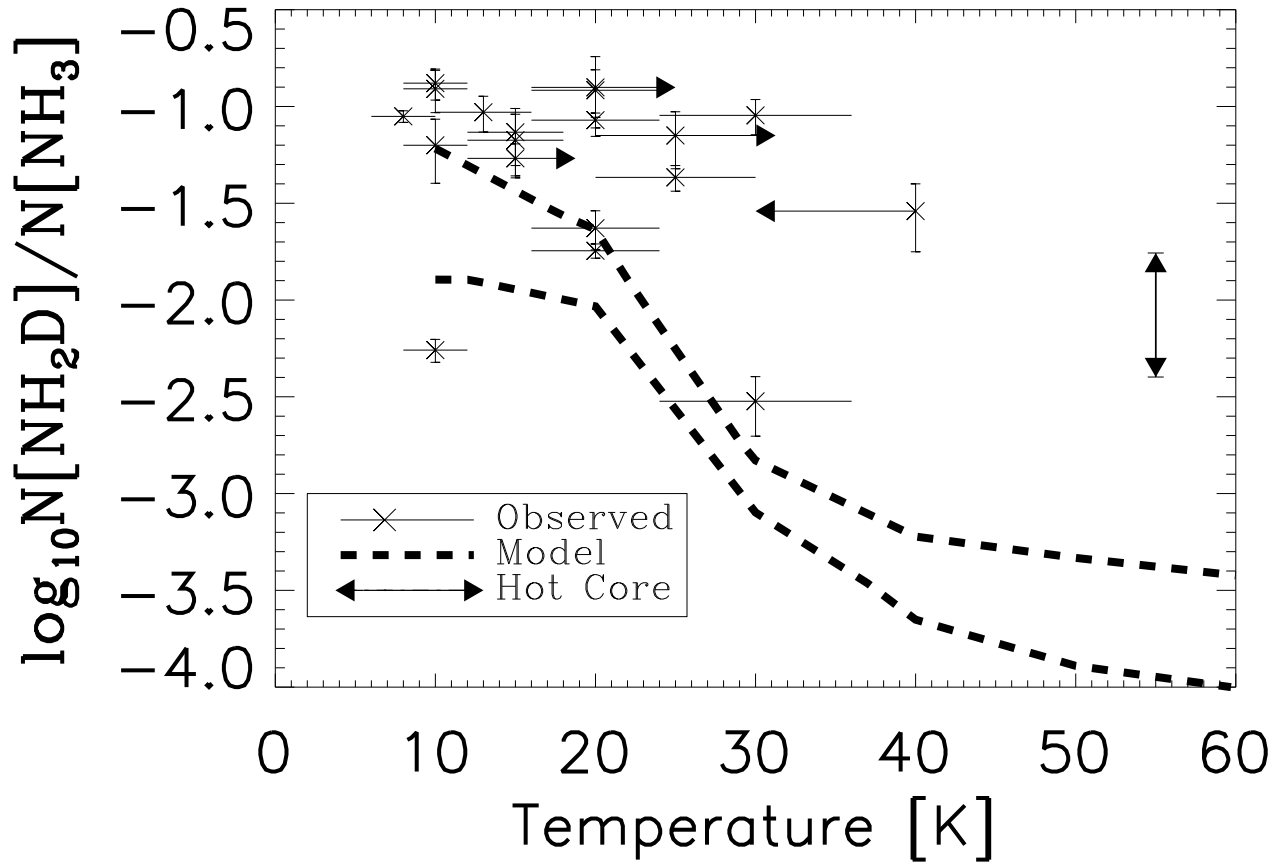}
\caption{Plot of ammonia fractionation versus temperature for sources
with single temperature \formal\ models (crosses).  We include the
predicted ammonia fractionation from the steady--state gas phase
models of Roberts \& Miller (2000) as upper and lower ranges marked
with dashed lines.  Additionally, we plot the range of values found
for hot cores by various authors ({\it N.B.} the temperature choice is
arbitrary) \label{NH3fracplot}}
\end{figure}

We plot in Figure \ref{NH3fracplot} the ammonia fractionation as a
function of source temperature listed in Table \ref{NH2Dratio}.
Observed values rise to $\sim 20$~K; sources warmer than this have a
flat \ratio\ ratio.  The range of \ratio\ for massive hot cores is
included as an arrow on the right side of the plot.  We include curves
(dashed--lines) based on the calculations of \citet{rm} that bracket
a range of values for ${\rm NH_2D/NH_3}$.  These model points are for
steady--state models at densities of $10^4$ and $10^5 {\rm cm^{-3}}$.
The initial conditions of their models include a cosmic D/H and
HD/${\rm H_2}$ ratio of $1.6\times 10^{-5}$ \citep{linsky1995}, a C/O
ratio of 0.4, and a cosmic ray ionization rate of $1.3\times 10^{-17}
{\rm s^{-1}}$.  The range of model values plotted in Figure
\ref{NH3fracplot} includes a depletion of C, N, and O abundances by
factors of 3 and 6.  Larger depletions and densities lead to higher ${\rm
NH_2D/NH_3}$ values.

The observational result suggests that Class 0 and dark core sources
have ${\rm NH_2D/NH_3}$ values set by the earlier conditions of the
parent cloud.  Thus, we observe the ``fossil remnants'' of gas--phase
chemistry when we study \ratio\ towards these sources.  The
theoretical predictions of \citet{rm}, however, appear to conflict
with the observed values, even for the coldest sources whose
formaldehyde models consist of single, isothermal zones.  The
time--dependence of ammonia fractionation suggests a way to resolve
this discrepancy.  \citet{rm} indicate that ${\rm NH_2D/NH_3}$ values
of $\sim 0.1$ are reached very quickly within a few $\times 10^3$ yrs,
for isothermal 10 K sources at moderate densities of $10^4 {\rm
cm^{-3}}$ and persists for roughly $10^{5.5}$ yrs.  In a pure
gas--phase model, this decreases to $\sim 10^{-2}$ after $10^{5.5}$
yrs, while for a model that includes active grain accretion the
fractionation rises above $10^{-1}$.  This suggests that all of the
sources except one (TMC1-CY) are at least $10^4$ yrs old.  However,
such age estimates are unconstrained.  Additional deuterium
fractionation ratios and abundances are necessary in order to fully
test the validity of such a ``chemical chronometer''.  One example
that may help calibrate such a clock is recent modeling of CO
depletion towards L1544 by \citep{caselli1999}.  Their dynamical and
chemical modeling of CO, ${\rm H^{13}CO^+}$, and ${\rm D^{13}CO^+}$
line profiles suggests an age of at least $10^4$ years.  They derive a
${\rm DCO^+}$/${\rm HCO^+}$ ratio of $0.12\pm 0.02$.  We find ${\rm
NH_2D/NH_3}=0.13\pm 0.02$ towards L1544 which, according to the model
of \citet{rm}, is consistent with the age derived by
\citep{caselli1999}.

The two primary Taurus positions, TMC1-CY and TMC1-${\rm NH_3}$ offers
a relative comparison of age estimates derived from deuterium
fractionation ratios.  \citet{pratap1997} mapped the spatial
distribution of 34 transitions of 14 molecules towards the Taurus
Molecular Cloud ridge, which includes these sources.  They argue that
variations in the emission from carbon--bearing and other molecules
result from both density and abundance variations.  \citet{pratap1997}
find especially striking abundance gradients for SO, ${\rm HC_3N}$,
and ${\rm CH_3CCH}$, which they explain with a small difference in the
chemical evolution time-scale between the northeast and southwest ends
of the cloud, or by a small change in the gas--phase C/O ratio.  Both
possibilities effectively reduce the carbon abundance, which would
explain the lower abundances of cyanopolyyenes in the northern regions
near the ammonia peak.  Comparisons of \ratio, \rdcop, \rhcn, and
${\rm N_2D^+/N_2H^+}$ with the chemical model of \citet{rm} satisfies
both of these requirements.  However, the age difference is preferred
since both ${\rm NH_2D}$ and ${\rm N_2H^+}$ are considered to be
largely derived from ${\rm N_2}$.  Molecular nitrogen is among the
least reactive of neutral species in the ISM, making it likely to be
abundant in the dense gas found in the more advanced stages of
protostellar evolution.

Two recent studies of \ratio\ by \citet{tine2000} and
\citet{saito2000} support and contradict, respectively, our conclusion
that gas--phase chemistry is responsible for the ammonia deuterium
ratios of L183 and TMC1-${\rm NH_3}$.  \citet{tine2000} compare their
observations with a simple--gas phase network based on the reaction
rates of \citet{umist91}.  They successfully reproduce their observed
abundances and deuterium fractionation ratios.
\citet{saito2000} on the other hand, compare ${\rm NH_2D}$ data for
several ammonia cores.  They invoke grain synthesis of deuterated
species even for their cold, star-less sources such as L183.
\citet{tine2000} observed L134N (1\arcmin\ NW of L183) and TMC1-N
(equivalent to TMC1-${\rm NH_3}$).  They experienced some ambiguity in
modeling the spectra with an LVG code and relied instead on LTE values
for determining \nh2d\ column densities, analysis, and fractionation:
${\rm N(L134N)}=\scnot{2}{14}$ and ${\rm N(TMC1-N)}=\scnot{1.3}{13}$;
${\rm X(L134N)}=\scnot{1.5}{-9}$ and ${\rm
X(TMC1-N)}=\scnot{1.2}{-10}$; ${\rm (L134N)}=0.18$ and ${\rm
(TMC1-N)}=0.02$.

We find N(\nh2d) equal to \scnot{3}{13} and \scnot{1}{13} \cm{-2}, one
and zero orders of mag different for L183 and TMC1-${\rm NH_3}$.
However, we have beam dilution between effects between our 12~m
observations and the 30~m survey of \citet{tine2000} and different
positions, for L183.  Our abundances, columns and fractionations for
TMC1-${\rm NH_3}$ are fairly similar.  \citet{tine2000} are able to
reproduce their observations with extensive gas--phase modeling of the
ammonia chemistry, with a few important caveats.  Moderate to
significant (factors of a few to 10) depletion of carbon and oxygen
(i.e. removal of neutral and atomic destroyers, and reduction of \xe)
will help achieve the high abundances and fractionations (0.02 and 0.1
for L134N and TMC1--N, respectively).  Our results for TMC1-${\rm
NH_3}$ and L183 are similar, though we find a factor of two smaller
deuterium fractionation towards L183.  Additionally, we have been able
to extend our analysis to other dark cores, which also indicates that
gas--phase chemistry is most important.  This suggests that grain
processes are unimportant for these low mass cores, at least for
ammonia formation and deuterium fractionation.

The results of \citet{saito2000} are discrepant with our view that
gas--phase synthesis dominates the formation of \nh2d.  They studied
\ratio\ for 16 ammonia cores from the \citet{bemy83} list and found
that mostly those with {\it IRAS} detections show elevated or
observable \nh2d\ abundances and \ammonia\ fractionation.
Furthermore, they find larger abundances than expected from gas--phase
networks for the kinetic temperatures of 4 sources.  They concluded
that ammonia is formed and deuterated on grains.  Several problems
exist with their analysis.  First, a notable exception to their
conclusions regarding grain--chemistry formation of \nh2d\ is L183,
where they find abundances and fractionations similar to our own.
Second, they use an LTE estimate of the excitation of \nh2d, which
tends to over-predict the column density.  This is because in an LTE
scenario, one requires a larger abundance (column density or optical
depth of \nh2d) to achieve the same line intensities than for non-LTE
methods.  Third and last, they did not integrate enough on all of
their sources, since TMC1-\ammonia\ has a clear detection by us and by
\citet{tine2000}.  Thus, it is also possible that for their dark cores
without infrared sources, the column density of \nh2d\ may be below
the detection limit of their survey.

Observational and theoretical comparison of L183 and TMC1-${\rm NH_3}$
provide important benchmarks for gas--phase synthesis of molecules
\citep{swade1987}.  L183 and TMC1--\ammonia\ possess similar values of
$n$, $T$, and size, but display very different chemical
characteristics.  Most chemical and dynamical models of L183 have
concentrated on the dearth of cyanopolyyenes and abundance of
sulferetted molecules.  Deuterium fraction may provide additional
constraints.  Ammonia fractionation is larger by a factor
of two in L183 when compared to TMC1--\ammonia.  Different initial
abundances between L183 and TMC1 have generally been ruled out by
their relative isolation and similar galactocentric distance.
TMC1--\ammonia\ should, as a whole, be older than L183 based on
time--evolution models of hydrocarbons; however the same cannot be
said for sulphur, which is observed to have lower abundances in TMC1
resulting in the conclusion that L183 is older.  \citet{swade1987}
points out that the conflicting chemical histories of L183 and
TMC1--\ammonia\ can be rectified in terms of larger depletions in
TMC1.  In order to achieve this, one requires a larger molecular
hydrogen {\it volume} density, which in cold gas leads to more easily
depleted material.  The slightly lower abundance of ammonia (which we
observe) is consistent with a larger depletion in TMC1--\ammonia.

\subsection{Sources with warm gas and more complex structure}

Our \formal\ modeling indicates that several sources (L1448IRS3,
NGC~1333 IRAS4A, NGC~1333 IRAS2, NGC~1333 IRAS7, S68~FIRS1, and S68N)
are well fitted with two distinct layers with different densities and
temperatures.  We do not know for a certainty where all of the \nh2d\
and \ammonia\ is located.  The reason for this discrepancy is the
inherent ambiguity in our source models.  We are able to arbitrarily
place the ammonia in either the warm or cold zones, and still
successfully reproduce the observed integrated intensity.  Therefore,
we know only what the {\it total} column density and column density
ratios are towards these sources.  The lack of knowledge about the
grain contributions to the observed deuterium fractionation makes an
accurate census difficult.  However, similarity between column density
ratios among Tables \ref{ammfracsingle} (cold, single layer) and
\ref{multilayer} (warm, multi-layer) does not suggest an extra warm gas
contribution is necessary.  Those profiles that show well--resolved
self--reversal provide a tool to break this degeneracy.  We
concentrate on IRAS4A, since it contains the most well resolved line
profiles, and is thus easiest to consider.  Higher spectral resolution
data are necessary for a less ambiguous estimate of the column density
distribution in the other self--absorbed sources.

\subsubsection{Foreground Absorption in IRAS4A}

An extended, cool envelope may explain the self--absorbed profile
observed towards IRAS4A.  We estimate a {\it lower} limit to the
self-absorbing layer of NGC~1333 IRAS4A.  We assume that the \nh2d\
absorbing layer is primarily cold gas near 25K located along the
line--of--sight in the central 0.5 \kms.  We estimate the lower limit
to the absorbing layer column density by assuming that the optical
depth is 1 and varying the temperature.  We find that the column
density must be at least a few times $10^{13}$\cm{-2} for temperatures
between 10 and 25 K (the temperature of the cold layer in our
micro-turbulent model).

This allows us to fix the column density in the outer layers of IRAS4
to {\it at least} \scnot{1}{13}\cm{-2} and estimate the total column
density with our micro-turbulent model by fitting the observed
integrated intensity listed in Table \ref{dammntottab}.  Using this as
an initial guess we estimate the column density towards IRAS4A in the
warm gas is at least a factor of 10 less than in the colder gas.
The ammonia data lacks the spectral resolution to permit a similar
analysis.  If we arbitrarily divide the \ammonia\ column density
evenly between the warm and cool layers ({\it i.e.} assume that
\ammonia\ itself is ubiquitous and well--mixed in the ISM), \ratio\
falls off by about a factor of 10.  This is roughly the expectation
from {\it gas--phase} chemical models of deuterium fractionation.
However, in addition to the lack of spectral resolution in the
\ammonia\ data, we are not resolving any spatial structure towards
IRAS4A.  High spatial resolution analysis are necessary to understand
the distributions of \nh2d\ and \ammonia.

\subsection{Grain Deuteration \label{grains}}

Grain enhancements may provide an important alternative source for
deuterium fractionation of ammonia for Class 0 sources.  Though we
suggest that the flat trend in \ratio\ for sources warmer than 20~K
indicates a gas--phase origin, it is important to discuss the general
scheme here.  Grain accretion time scales in dense cores are of order
the collapse time scale, $\lsim 10^6$yrs.  The larger mass of
deuterium allows it to stick longer to a surface, and bind more
strongly than hydrogen with other atoms.  Observations of ${\rm
D_2CO}/{\rm H_2CO}$ \citep{turner1990,castets1999}, ${\rm HDO}/{\rm
H_2O}$ \citep{jacq1990}, and ${\rm CH_3OD}/{\rm CH_3OH}$
\citep{mauersberger1988}, underscore the importance of grain--based
molecular formation and fractionation, although more recent gas--phase
synthesis models (e.g. Roberts \& Millar 2000) suggest otherwise.

A grain origin for ${\rm NH_2D}$ is suggested based on the similarity
of the fractional abundances of \nh2d\ in TMC1 versus the Orion Hot
Core.  A review of the abundances using standard methods is appropriate
at this point.  \citet{walmsley1987} find that the \nh2d\ emission is
optically thin in the Orion--KL region.  They obtain
N(\nh2d)=\scnot{1.54\pm 0.4}{14}.  Using the dust continuum
measurements of \citet{masson1985} to estimate the \nh2d\ fractional
abundance, one finds $X[{\rm NH_2D}]$=\scnot{1.01-6.21}{-10}.  Using
the ${\rm C^{18}O}$ value from \citet{ungerechts1997} and the
conversion factor for carbon monoxide to molecular hydrogen developed
by \citet{frerking1982}, ${\rm N(H_2)}$ is a factor of 10 smaller than
the dust value.  Therefore, the abundance of deuterated ammonia is
equivalently larger by ten.  Towards TMC1, only ${\rm C^{18}O}$ data
is available, and we find ${\rm NH_2D}/{\rm H_2}=\scnot{4.91\pm
1.63}{-10}$, quite similar to the Orion Hot Core.  Abundance
comparisons, in general, are problematic, especially between
physically different regions such as TMC1 and the hot core associated
with IRc2.  CO to molecular hydrogen conversion factors and
line-of-sight complexity towards a source are difficult to properly
estimate.  Therefore, abundance studies should be approached with some
caution\footnote{See the discussion by \citet{mundymcmullin1996}.}.

An important distinction between our source list and hot cores used in
studies such as \citet{walmsley1987} and \citet{jacq1990} is the
temperature.  Desorption mechanisms will be largely ineffective at
removing any depleted or surface-chemistry products in dark clouds,
unless low-temperature, non-thermal desorption such as \hydrogen\
formation on grains (which releases $\sim 4.5$ eV of energy) occurs
rapidly.  This will only occur for low density regions, where
\hydrogen\ formation theoretically prevents the condensation of CO and
water on small grains and PAHs \citep{duleywilliams1993}.  Even for
the more evolved Class 0 objects included here, grain chemistry can be
excluded because of the sublimation temperature of ices.  Grain
mantles are dominated by water ice (e.g. Ehrenfreund \& Charnley
2000), whose sublimation temperature is 90K.  Since our sources are
all cooler than this, grain sublimation is not likely.  In fact,
thermal grain desorption at low temperatures and densities above
$10^5$~\cm{-3} proves quite inefficient, even in the most optimistic
models where tunneling on small grains occurs, reaction
exothermicities exceed binding energies, and H/${\rm H_2}$ is large
enough to quickly saturate C, O, and N atoms.  We discuss several
other grain desorption mechanisms below, under the assumption that
surface deuterium enhancements may contribute to the observed \ratio.
These include: thermal, cosmic-ray induced grain heating,
photo-desorption, collisional (grain-on-grain violence).  Impulse
heating of grains (temporarily increasing the temperature of a single
grain above the threshold for thermal desorption) is achieved with
X-rays and cosmic rays.  \citet{leger1985} investigated this affect in
regions of varying density and visual extinction.  For
$n<10^4$~\cm{-3} and $A_V<5$, desorption efficiently removes CO and
even water--ammonia ice mantles quite readily.  However, for more
dense regions, the water-ammonia lattice will remain, since for
$A_V\gsim$few UV photons cannot penetrate them.
\citet{wootten1982,butner1995,williams1998} have shown that X--ray
contributions appear to be minimal since the electron fraction
measured via \rhcop\ varies little for sources with and without stars.
Therefore, our sources would have little cosmic--ray desorbed input
from abundances of ammonia either stored or produced on grains.

%%
%% Conclusions
%%
\section{Conclusions \label{Conclusions}}

We observed \nh2d\ hyperfine transitions towards 32 protostellar and
pre--stellar sources with similar spatial resolution.  We find nearly
a 70\% detection rate for the \nh2d\ lines at 85.9~GHz.  The observed
abundances of \nh2d\, when compared to single--dish \ammonia\
observations of similar beam--width, indicate that the deuterium
fractionation is large, $10^{-3}\lsim$\ratio $\lsim 10^{-1}$.  The
observed \ratio\ values generally exceed or equal those seen in hot
core regions as well as in warm, embedded condensations in otherwise
low luminosity sources with large, cold envelopes.  Sources with
$T_K\lsim 20$~K follow gas--phase predictions for \ratio; for
$T_K>20$~K the trend flattens, indicating that Class 0 sources have
not begun to destroy \nh2d.  ${\rm NH_2D/NH_3}$ ratios reflect the
``fossil'' remnants of gas--phase synthesis.  We additionally conclude
that grain formation of deuterated ammonia is not necessary to explain
${\rm NH_2D/NH_3}$.

Some evidence exists that dynamical ages of protostellar sources can
be derived from comparisons of deuterium fractionation and chemical
models of star forming cores.  However, this notion is currently
rather speculative.  We are able to address {\it relative}
evolutionary differences between sources such as TMC1-CY, TMC1-${\rm
NH_3}$, and L183 successfully.  

Self--reversed profiles indicate that the overall story for \nh2d\ is
far more complicated, however.  Indeed, future careful modeling and
observations will attempt to address the coupled influences of
chemistry and dynamical evolution in IRAS4.  Nonetheless, \nh2d\
provides important insights into deuteration in the ISM.  Deuterium
fractionation not only varies among sources of different physical
attributes, but also among different molecules.  Thus, a full census
of deuterium fractionation seems necessary in order to test
astrochemical models.  Furthermore, with deuterated molecules now
providing information on the collapse regions of protostars, new and
potentially powerful tools for examining the history of core collapse
via chemical evolution models can provide checks on more traditional
methods.

\acknowledgements

We thank Barry Turner for many helpful comments and discussions.  RYS
thanks NRAO for support through the Junior Research Associate program.
RYS also expresses profound appreciation to the NRAO Tucson/12--meter
staff for their aid through his thesis work.  Research support at the
Laboratory for Astronomical Imaging is supported by NSF grant AST
99-81363 and by the University of Illinois.

%%
%% References
%%

%%
%% Figures with captions
%%
%\clearpage

%\clearpage

%\clearpage

\clearpage
\begin{figure}
\plotone{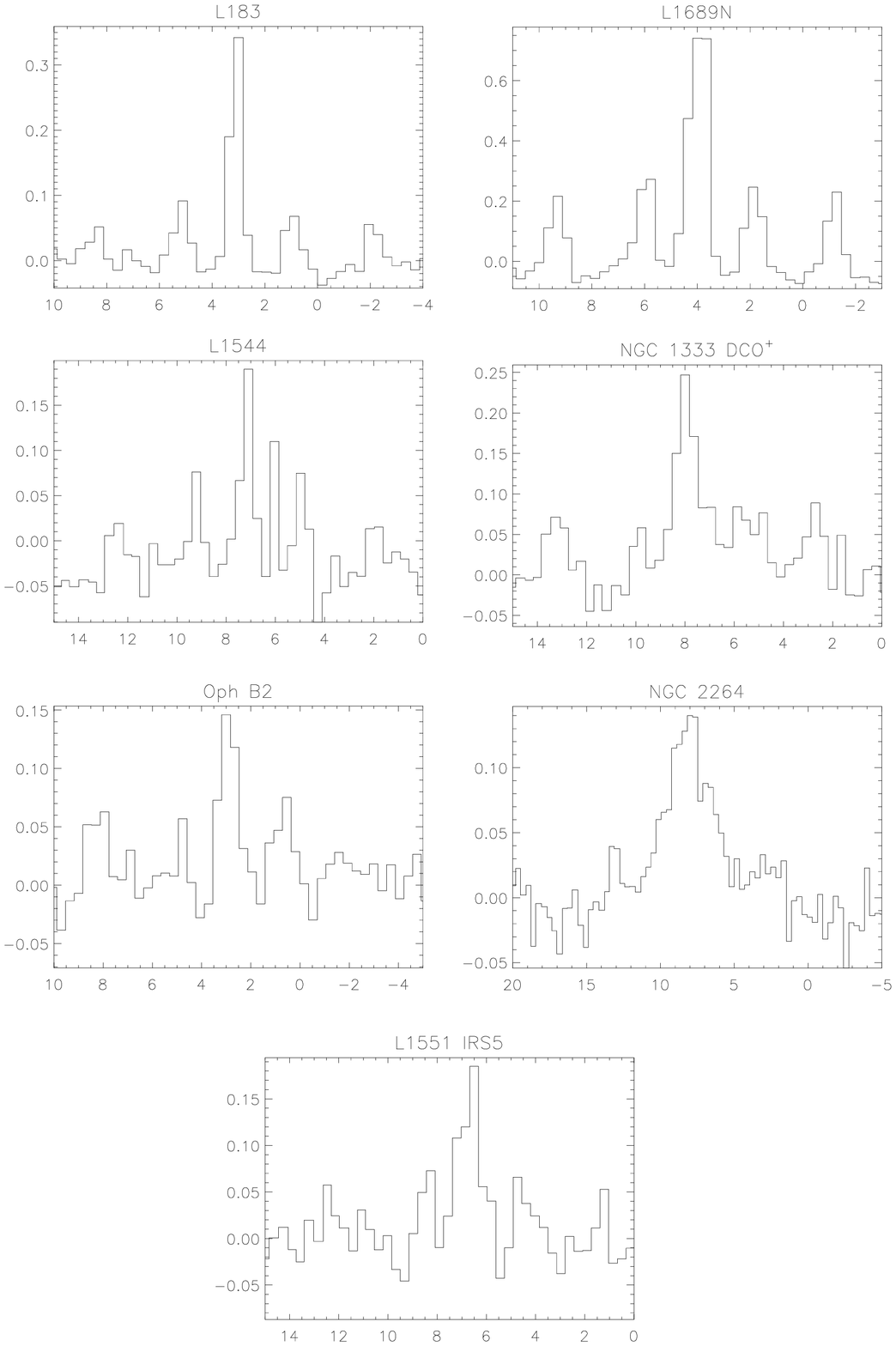}
\caption{110.1 GHz \nh2d\ spectra.  The spectral resolution
is 0.35 \kms/channel.  A linear baseline is removed from the data,
revealing the hyperfine components.  \label{fig:plot110_1}}
\end{figure}

%\clearpage

%Figure of L1689N in the 1-0 Fundamental of NH2D
%\clearpage
\begin{figure}
\plotone{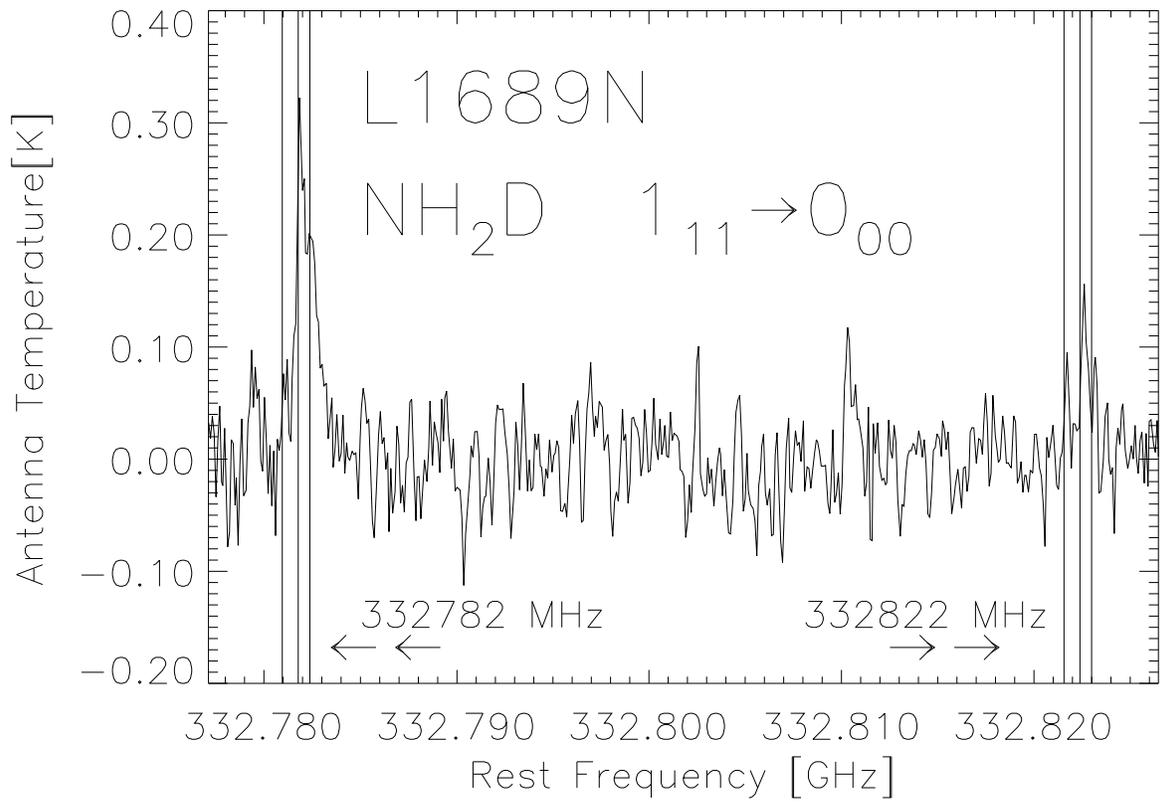}
\caption{Plot of the 333 GHz $\jeq{1_{11}}{0_{00}}$ fundamental transition
\nh2d\ observed with the Caltech Submillimeter Observatory.  The
spectral resolution is 0.04 \kms\ for this 1.5 hour spectrum. Both
sets of lines near 332.820 and 332.782 GHz are heavily blended.  The
integrated intensity is consistent with an LTE distribution.
\label{fig:L1689Nfundy}}
\end{figure}

\clearpage
%Figure of Ammonia fractionation versus temperature
%	Includes results of Millar, Bennet, and Herbst (1989)
% used to use modelvsobs.eps

%%
%% Tables
%%
%\clearpage
%% Table of frequencies for NH2D
\begin{deluxetable}{cccc}
\tabletypesize{\footnotesize}
\tablecolumns{4}
\tablewidth{0pt}
\tablecaption{Deutero--Ammonia Transition Frequencies \label{nh2dfreqtable}} 
\tablehead{
\colhead{Transition} &
\colhead{$F\arcmin - F\arcsec$} & 
\colhead{Intensity} &
\colhead{Frequency (GHz)\tablenotemark{\dag}}}
\startdata
                       & 0--1 & 0.111 & 85.9247829 \\
		       & 2--1 & 0.139 & 85.9257031 \\
$\jeq{1_{11}}{1_{01}}$ & 2--2 & 0.417 & 85.9262703 \\
$\jeq{1_{11}}{1_{01}}$ & 1--1 & 0.083 & 85.9263165 \\
		       & 1--2 & 0.139 & 85.9268837 \\
 		       & 1--0 & 0.111 & 85.9277345 \\ 
\tableline
 		       & 0--1 & 0.111 & 332.7809447 \\
$\jeq{1_{01}}{0_{00}}$ & 2--1 & 0.556 & 332.7817955 \\
		       & 1--1 & 0.333 & 332.7823627 \\
\tableline
\tableline
		       & 0--1 &	0.111 & 110.152084 \\
		       & 2--1 & 0.139 & 110.152995 \\
$\jeq{1_{11}}{1_{01}}$ & 2--2 & 0.417 & 110.153599 \\
$\jeq{1_{11}}{1_{01}}$ & 1--1 & 0.083 & 110.153599 \\
		       & 1--2 & 0.139 & 110.154222 \\
		       & 1--0 & 0.111 & 110.155053 \\
\tableline
		       & 0--1 & 0.111 & 332.8215595 \\
$\jeq{1_{01}}{0_{00}}$ & 2--1 & 0.556 & 332.8224149 \\
		       & 1--1 & 0.333 & 332.8229853 \\
\enddata
\tablenotetext{\dag}{Tin\'{e} et al. 2000; Townes \& Schawlow 1955}
\end{deluxetable}

%\clearpage
%% Table of sources
\begin{center}
\begin{deluxetable}{lccc}
\tabletypesize{\footnotesize}
\tablecolumns{4}
\tablewidth{0pt}
\tablecaption{Source List\label{srclist}}
\tablehead{
\colhead{Source} & 
\colhead{{\it R. A.} (1950)} & 
\colhead{{\it DEC} (1950)} &
\colhead{$V_{LSR}$} \\ 
\colhead{} & 
\colhead{\it h\ \ m\ \ s} & 
\colhead{\arcdeg\ \ \arcmin\ \ \arcsec} & 
\colhead{\kms}}
\startdata
L1448~IRS2 &03 22 17.9  &30 34 41 &   4.0\\
L1448~NW   &03 22 31.1  &30 35 14 &   5.4\\
L1448~IRS3 &03 22 31.9  &30 34 45 &   4.7\\
L1448~C    &03 22 34.4  &30 33 35 &   5.4\\
NGC~1333~\dcop     &03 26 3.60  &31 04 42 &   7.0\\ 
NGC~1333 IRAS7     &03 26 6.90  &31 08 28 &   7.0\\ 
NGC~1333~IRAS2     &03 25 52.6  &31 04 30 &   7.0\\
NGC~1333~IRAS4A    &03 26 4.78  &31 03 14 &   7.0\\ 
NGC~1333~IRAS4B    &03 26 7.00  &31 02 52 &   7.0\\
NGC~1333~IRAS4C    &03 26 8.10  &31 03 37 &   7.0\\ 
IRAS~03282         &03 28 15.2  &30 35 14 &   7.0\\
IRAS~03367--IC~348 &03 36 47.1  &31 47 29 &   7.0\\
L1551      &04 28 40.2  &18 01 42 &   6.5\\
TMC1~AM    &04 38 19.5  &25 42 29 &   6.0\\
TMC1~CY    &04 38 38.0  &25 35 45 &   6.0\\ 
L1512     &05 00 54.5  &32 40 00 &   7.9\\ 
L1544     &05 01 15.0  &25 07 00 &   7.1\\
OMC2      &05 32 58.0  &-05 12 11&   11.0\\
HH1       &05 33 52.0  &-06 47 09&   10.0\\
NGC~2264  &06 38 24.9  &09 32 29 &   8.0\\
L183      &15 51 30.0 &-02 43 31 &   3.0\\ 
VLA1623   &16 23 25.0  &-24 17 47&   4.0\\
OPHB1     &16 24 09.0 &-24 22 49 &   3.0\\
OPHB2     &16 24 26.3 &-24 19 49 &   3.0\\
IRAS 16293    &16 29 21.0 &-24 22 16 &   4.1\\ 
L1689N    &16 29 27.6 &-24 22 08 &   4.0\\ 
L63       &16 47 17.0 &-18 00 00 &   6.0\\ 
S68FIR    &18 27 17.5  &01 13 23 &   8.0\\ 
S68N      &18 27 15.9  &01 14 49 &   9.0\\ 
\enddata
\end{deluxetable}
\end{center}

%\clearpage
%% Fits to ammonia and d-ammonia data
\begin{deluxetable}{lcccccl}
\tabletypesize{\footnotesize}
\tablecolumns{7}
\tablewidth{0pt}
\tablecaption{Deutero--Ammonia Microturbulent Column Density Fits
\label{dammntottab}} 
\tablehead{
\colhead{Source} &
\colhead{$\Delta$V} & 
\colhead{$T_r^*\Delta V$} &
\colhead{$T_{K}$\tablenotemark{\dag}} & 
\colhead{$n{\rm (H_2)}$\tablenotemark{\dag}} & 
\colhead{N(\nh2d)} & 
\colhead{Temp.} \\
\colhead{} & 
\colhead{\kms} & 
\colhead{K \kms} & 
\colhead{K} &
\colhead{$10^6~{\rm cm}^{-3}$} & 
\colhead{$10^{12}~{\rm cm}^{-2}$} &
\colhead{Ref.}}
\startdata	
L1448~IRS2 & 0.49 & $0.84\pm 0.07$ & 20 & 0.1 &
$7.60\pm 0.61$ & O'Linger et al. 1999 \\ 
L1448~NW& 1.38&$3.01\pm 0.60$ & 30 & 0.1&
$36.0\pm 7.20$ & Barsony et al. 1998 \\ 
L1448~IRS3 & 1.10 & $1.39\pm 0.09$ & 20--50 & 0.1-5 &
$8.14\pm 0.55$ & This paper \\  
L1448~C& 0.93& $0.81\pm 0.26$ & $40\pm 10$& 1 &
$9.30\pm 2.96$ & This paper \\ 
NGC~1333~\dcop& 2.60& $3.77\pm 0.96$ & 20 & 0.6 &
$45.0\pm 11.5$ & This paper\\ 
NGC~1333~IRAS2 & 1.31 & $0.46\pm 0.09$ & 20--85 & 0.4-1.26 &
$2.62\pm 0.49$ & Wootten et al. 1999 \\
NGC~1333~IRAS7 & 2.31 & $2.30\pm 0.46$  & 40 & 0.3 &
$2.60\pm 0.52$ & Lefloch et al. 1998 \\ 
NGC~1333~IRAS4A\tablenotemark{[1]}& 1.50 & $3.98\pm 1.19$ & 25--50 & 0.7-2
&$21.8\pm 6.50$ & Wootten et al. 1999 \\ 
NGC~1333~IRAS4C& 1.10& $1.87\pm 0.11$& 15& 0.1 &
$15.4\pm 0.90$ & This paper \\ 
IRAS~03282 & 0.90& $0.76\pm 0.23$& 20& 0.1 &
$7.00\pm 2.12$ & Wootten et al. 1999 \\ 
L1551 & 1.35 & $1.04\pm 0.11$& 50 & 0.1 &$22.5\pm
2.40$ & Wootten et al. 1999 \\ 
TMC1~AM& 1.12& $0.44\pm 0.12$& 10& 0.1 & $9.77\pm
2.57$ & This paper \\ 
L1544& 0.96& $2.76\pm 0.12$& 10& 0.3 & $26.0\pm
1.10$ & This paper \\
OMC2 &     1.66& $2.55\pm 0.37$& 20 & 0.1 & $220.0\pm
32.0$ & Batrla et al. 1983\\ 
HH1  &     1.68& $1.04\pm 0.33$& 15 & 0.1 & $100.0\pm
32.0$ & Martin--Pint. (1987) \\ 
NGC~2264 & 3.10 & $0.41\pm 0.08$ & 25 & 0.1 & $34.4\pm
5.20$ & de Bois. et al. 1996 \\ 
L183 & 0.30\tablenotemark{[2]} & $2.40\pm 0.42$& 10& 0.01 &$30.0\pm
11.1$ & This paper \\ 
VLA~1623 & 1.04& $1.92\pm 0.17$& 20& 0.1 & $17.0\pm
1.50$ & AWB93 \\ 
Oph~B2& 1.33& $1.90\pm 0.40$ & 13 & 0.1 & $9.00\pm
1.89$ & Lefloch et al. 1998 \\ 
IRAS~16293& 0.81& $1.24\pm 0.12$ & 30 & 0.6 &$4.50\pm
0.30$ & van Dish. et al. 1995 \tablenotemark{\ddag}\\   
L1689N & 1.22 & $6.46\pm 0.45$ & 8 & 0.1 & $160.0\pm
11.0$ & Wootten (pc) \\ 
S68~FIR &1.58 & $1.23\pm 0.43$ & 20--100 & 0.3-1.26 &
$34.0\pm 12.00$ & McMullin et al. 1999 \\ 
S68N & 1.46& $1.32\pm 0.15$& 15--75& 0.3-2 &
$7.82\pm 0.87$ & McMullin et al. 1999 \\ 
\enddata
\tablenotetext{\dag}{Sources with two values list the \formal\ fits
for density and temperature in two layer models}
\tablenotetext{\ddag}{From the fit to their cold, extended layer}
\tablenotetext{[1]}{The error listed contains an estimate to the error 
in the linewidth of 0.3 \kms.}
\tablenotetext{[2]}{Using line fits from high resolution 30~kHz
spectrum.}  
\end{deluxetable}

\clearpage
% upper limits to NH2D
\singlespace
\begin{center}
\begin{deluxetable}{lcc}
\tabletypesize{\footnotesize}
\tablecolumns{3}
\tablewidth{0pt}
\tablecaption{\nh2d\ $5\sigma$ Upper Limits \label{upperlims}}
\tablehead{
\colhead{Source} &
\colhead{$T_r^*\Delta V$} & 
\colhead{N(\nh2d)}\\ 
\colhead{} & 
\colhead{K \kms} &
\colhead{${\rm cm}^{-2}$}}
\startdata
IC~348 & 0.094 & \scnot{2}{12}\\
%L1529  & 0.340 & \scnot{5}{12}\\
L1512  & 0.024 & \scnot{1}{12}\\
TMC1~CY& 0.007 & \scnot{5}{11}\\
%Oph~A  & 0.045 & \scnot{1}{12}\\
Oph~B1  & 0.045 & \scnot{1}{12}\\
L63    & 0.060 & \scnot{1.5}{12}\\
\enddata
\tablenotetext{\ddag}{{\it N.B.} Upper limits from $T_K=20K$ and $n{\rm
(H_2)}=$\scnot{1}{5}}
\end{deluxetable}
\end{center}

%\clearpage
% NH3 fits
\begin{center}
\begin{deluxetable}{lcccccl}
\tabletypesize{\footnotesize}
\tablecolumns{7}
\tablewidth{0pt}
\tablecaption{Ammonia Microturbulent Column Density Fits\label{ammntottab}} 
\tablehead{
\colhead{Source\tablenotemark{\dag}} &
\colhead{$\Delta$V} &
\colhead{$T_r^*\Delta V$} & 
\colhead{$T_{K}$\tablenotemark{\ddag}} &
\colhead{$n({\rm H_2})$\tablenotemark{\dag}} & 
\colhead{N({\ammonia})} & 
\colhead{Ref.} \\
\colhead{} & 
\colhead{\kms} & 
\colhead{K \kms} &
\colhead{K} & 
\colhead{$10^6$~\cm{-3}} &
\colhead{$10^{14}$~\cm{-2}} &
\colhead{}}
\startdata
L1448~IRS2&0.87&$6.21\pm 1.32$&  20 & 1 & $3.23\pm
0.69$ & This paper\\
L1448~NW & & & & & $2.00\pm 0.20$ & Barsony et al. 1998\\ 
L1448~IRS3& 1.22& $14.74\pm 0.74$& 20--50 (30) & 0.1-5 &
$4.53\pm 0.23$ & This paper\\ 
L1448~C& 1.37& $11.07\pm 2.35$& $40\pm 10$& 1 &
$3.23\pm 0.69$ & This paper\\
NGC~1333~\dcop& 1.56& $15.07\pm 1.56$& $<30$ (38)& 0.6&
$3.71\pm 0.38$ & This paper\\ 
NGC~1333~IRAS4A& 1.52& $12.09\pm 1.61$& 25--50 & 0.7-2
& $3.08\pm 0.41$ & Blake et al. 1995\\ 
NGC~1333~IRAS4C& 1.44& $9.73\pm 3.50$& 25 (40)& 0.1
&$2.30\pm 0.82$& This paper\\ 
IRAS~03282\tablenotemark{\dag} & & & & & $10.0\pm 2.0$ & BMP91\\ 
IC~348\tablenotemark{\dag} & & & & & $2.50\pm 1.50$ & BGK87 \\
%L1529 & & & & & & \\
L1551& 1.27& 0.78& 5.33& 0.1 & $2.60\pm 0.26$ & This paper\\
TMC1~\ammonia& 0.81& $5.48\pm 1.37$& 20& 0.1 &$1.55\pm
0.39$ & Pratap et al. 1997\\ 
TMC1~CY& 0.72& $3.34\pm 0.45$& 20& 0.1 & $0.91\pm
0.12$ & Pratap et al. 1997\\ 
L1512\tablenotemark{\dag} & & & & & $7.94$& BM83 \\
L1544 & 0.73& $5.04\pm 0.89$& 10& 0.3 & $1.97\pm
0.35$ & This paper\\ 
OMC2\tablenotemark{\dag} & & & & & $75.9\pm 8.0$ & CW94 \\
HH1\tablenotemark{\dag} & & & & & $5.00\pm 0.38$ & MP87 \\ 
NGC~2264\tablenotemark{\dag} & & & & & $8.00$& Kr\"ugel
et al. 1996 \\ 
L183& 0.73& $7.52\pm 1.45$& 10& 0.01 & $5.19\pm
0.91$ & Olberg et al. 1983 \\
VLA1623\tablenotemark{\dag} & & & & & $2.00$ & Wootten et al. 1999\\
%Oph~A & & & 45 & & & \\
Oph~B1\tablenotemark{\dag} & & & 19 & & $3.45\pm 1.75$ &
MP83,ZBW84 \\ 
Oph~B2 & 1.20 & 4.08 & 13 & 0.1 & $0.96$ & MP83 \\ 
IRAS~16293\tablenotemark{\dag} & & & & & $15.0\pm 5.0$ &
Mundy et al. 1995\\ 
L1689N& 1.00 & 0.55 & 8 & 1 & $18.0$ & Wootten (pc)\\
L63\tablenotemark{\dag} & & & & & $10.0$ & BM83 \\
S68~FIR&1.25& $8.83\pm 2.33$ & 20--100 & 0.3-1.3 &
$2.71\pm 0.72$ & This paper\\  
S68N&1.24& $5.17\pm 0.77$ & 15--75 & 0.3-2 & $1.45\pm 
0.22$ & This paper\\ 
\enddata
\tablenotetext{\dag}{\ammonia\ column densities for Sources without
listed line parameters are taken from the noted references.}
\tablenotetext{\ddag}{Sources with two values list the density and
temperature in two layer models} 
\tablenotetext{1}{MP83=\citet{pintado83}}
\tablenotetext{2}{ZBW84=\citet{zeng84}}
\tablenotetext{3}{MP87=\citet{pintado87}}
\tablenotetext{4}{BM83=\citet{bemy83}}
\tablenotetext{5}{CW94=\citet{cesaroni94}}
\tablenotetext{6}{BGK87=\citet{bachiller87}}
\tablenotetext{7}{BMP91=\citet{bachiller91}}
\end{deluxetable}
\end{center}

%\clearpage
% Single Layer models
\begin{center}
\begin{deluxetable}{lccc}
\tabletypesize{\footnotesize}
\tablecolumns{4}
\tablewidth{0pt}
\tablecaption{Ammonia Fractionation for Single Temperature and density
models \label{ammfracsingle}}
\tablehead{
\colhead{Source} & 
\colhead{$T_{K}$} & 
\colhead{$n({\rm H_2})$} & 
\colhead{\ratio} \\ 
\colhead{} & 
\colhead{K} & 
\colhead{\cm{-3}} &
\colhead{}}
\startdata
L1448~IRS2 & 20 & \scnot{1}{6} & $0.024\pm 0.005$ \\
L1448~NW   & 30 & \scnot{1}{6} & $0.09\pm 0.02$ \\
L1448~C & $40^{+30}_{-10}$ & \scnot{1}{6} & $0.029\pm 0.006$ \\
NGC~1333~\dcop & $<30$ & \scnot{1}{6} & $0.10\pm 0.03$ \\
NGC~1333~IRAS4C & 25 & \scnot{1}{6} & $0.067\pm 0.013$ \\
IRAS~03282 & 20 & \scnot{1}{5} & $0.007\pm 0.003$ \\
L1551 & 50 & \scnot{7}{5} & $0.087\pm 0.013$ \\
TMC1~\ammonia & 20 & \scnot{1}{5} & $0.06\pm 0.02$ \\
L1544 & 10 & \scnot{1}{5} & $0.13\pm 0.02$ \\
OMC~2\tablenotemark{2} & 20 & \scnot{1}{5} & $0.029\pm 0.006$\\
HH1 & 15 & \scnot{1}{5} & $0.074\pm 0.024$\\
NGC~2264\tablenotemark{1} & 25 & \scnot{1}{5} & $0.043\pm 0.007$ \\
L183 & 10 & \scnot{1}{4} & $0.058\pm 0.010$ \\
VLA~1623 & 20 & \scnot{1}{6} & $0.085\pm 0.008$ \\
Oph~B2\tablenotemark{3} & 13 & \scnot{1}{5} & $0.090\pm 0.019$ \\
IRAS~16293 & 30 & \scnot{6}{5} & $0.0028\pm 0.001$ \\
L1689N & 8 & \scnot{1}{5} & $0.089\pm 0.006$ \\
\enddata
\tablenotetext{1}{\ammonia\ data from Kr$\ddot{\rm u}$gel et al. (1996).}
\tablenotetext{2}{\ammonia\ data from Cesaroni \& Wilson (1994)}
\tablenotetext{3}{Martin--Pintado et al. 1983}
\end{deluxetable}
\end{center}

%\clearpage
% Multiple layer models
\begin{center}
\begin{deluxetable}{lrrr}
\tabletypesize{\footnotesize}
\tablecolumns{4}
\tablewidth{0pt}
\tablecaption{Multi--layer Models of Ammonia Deuteration\label{multilayer}}
\tablehead{
\colhead{Source} & 
\colhead{$n$} & 
\colhead{$T$} &
\colhead{\ratio\tablenotemark{(1)}} \\    
\colhead{} & 
\colhead{\cm{-3}} & 
\colhead{K} & 
\colhead{}}
\startdata
NGC~1333 IRAS4A & \scnot{7}{5} & 25& 0.07\\
                & \scnot{2}{6} & 50& \\
S68~FIR         & \scnot{3}{5} & 20& 0.13\\
                & \scnot{1.26}{6} & 100 & \\
S68N            & \scnot{3}{5} & 15& 0.05\\
                & \scnot{2}{6} & 75& \\
L1448~IRS3      & \scnot{1}{5} & 20& 0.02\\
                & \scnot{5}{6} & 50& \\
\enddata
\tablenotetext{(1)}{Summed over both layers.}
\end{deluxetable}
\end{center}

\clearpage
%% Ortho-Para ratio for NH2D
\begin{deluxetable}{lcccccc}
\tabletypesize{\footnotesize}
\tablecolumns{5}
\tablewidth{0pt}
\tablecaption{\nh2d\ Ortho--Para Ratio \label{orthopara}}
\tablehead{
\colhead{Source} & 
\colhead{\anttemp(110 GHz)} &
\colhead{\anttemp(86 GHz)} &
\colhead{$\frac{\anttemp(86 GHz)}{\anttemp(110 GHz)}$} & 
\colhead{$\frac{\rm N(86 GHz)}{\rm N(110 GHz)}$} \\ 
\colhead{} &
\colhead{K~\kms} &
\colhead{K~\kms}& 
\colhead{} &
\colhead{}}
\startdata
   N1333&$0.598\pm 0.142$&$2.447\pm 0.623$&$4.09\pm
1.43$& $4.11\pm 1.30$\\
   L1544&$0.295\pm 0.144$&$1.792\pm 0.075$&$6.08\pm
2.98$& $6.10\pm 3.02$\\
    OMC2&$0.90\pm 0.13$&$1.66\pm 0.24$&$1.84\pm
0.38$& $1.84\pm 0.90$\\
     HH1&$0.41\pm 0.15$&$0.68\pm 0.21$&$1.66\pm
0.79$& $1.66\pm 0.92$\\
    L183&$0.40\pm 0.06$&$1.56\pm 0.27$&$3.95\pm
0.92$& $3.96\pm 0.92$\\
   OPHB2&$0.42\pm 0.11$&$1.23\pm 0.26$&$2.94\pm
1.01$& $2.95\pm 1.20$\\
  L1689N&$1.59\pm 0.21$&$4.19\pm 0.29$&$2.63\pm
0.39$& $2.64\pm 0.45$\\
   L1551&$0.30\pm 0.24$&$0.69\pm 0.07$&$2.31\pm
1.85$& $2.31\pm 1.80$\\
\enddata
\end{deluxetable}

%\clearpage
%% Fractionation values
\singlespace
\begin{center}
\begin{deluxetable}{lcccc}
\tabletypesize{\footnotesize}
\tablecolumns{5}
\tablewidth{0pt}
\tablecaption{Ammonia Deuteratium Fractionation \label{NH2Dratio}}
\tablehead{
\colhead{Source} & 
\colhead{T$_K$[K]} & 
\colhead{\ratio} &
\colhead{\rdcop \tablenotemark{\dag}} &
\colhead{\rhcn\tablenotemark{\dag}}}
\startdata
L1448~IRS2 & 20 & $0.024\pm 0.005$ & \nodata& \nodata\\
L1448~NW   & 30 & $0.09\pm 0.02$   & \nodata& \nodata\\
L1448~IRS3 & 20 & $0.018\pm 0.001$ & \nodata& \nodata\\
L1448~C    & 40 & $0.029\pm0.011$  & \nodata& \nodata\\ 
NGC~1333~\dcop & 30 & $0.10\pm 0.03$ & 0.024\tablenotemark{1}& \nodata\\
NGC~1333 IRAS4A& 25--50 & $0.071\pm 0.023$ & \nodata& \nodata\\
NGC~1333 IRAS4C& 15 & $0.067\pm 0.024$ & \nodata& \nodata\\
IRAS~03282 & 20 & $0.007\pm 0.003$ & \nodata& \nodata\\
IC~348 & 20 & $<0.008$ & \nodata& \nodata\\
%L1529 & 15 & & \nodata& \nodata\\
L1551 & 50 & $0.087\pm 0.013$ & 0.035\tablenotemark{1}&
$0.016\pm0.001$\tablenotemark{3}\\
TMC1~AM &10 & $0.06\pm 0.02$ & 0.027\tablenotemark{2}&
0.022\tablenotemark{3}\\
TMC1~CY& 10 & $<0.006$& 0.00004& \nodata\\
L1512 & 20 & $<0.006$ & \nodata& \nodata\\
L1544& 10 & $0.13\pm 0.02$ & $0.12\pm 0.02$ & \nodata\\
OMC2& 20 & $0.029\pm 0.006$ & \nodata & 0.01\tablenotemark{3}\\
HH1 & 15 & $0.074\pm 0.024$ & \nodata & \nodata\\
NGC~2264& 25 & $0.043\pm 0.007$ & 0.017\tablenotemark{1}& \nodata\\
L183& 10 & $0.058\pm 0.010$ & 0.07\tablenotemark{1}& \nodata\\
VLA1623 & 20 & $0.085\pm 0.008$ & \nodata& \nodata\\
%Oph~A & 45 & & \nodata& \nodata\\
Oph~B1 & 19 & $<0.003$ & \nodata& \nodata\\
OPH~B2&13 & $0.09\pm 0.02$ & \nodata& 0.013\\
IRAS~16293 & 30 & $0.003\pm 0.001$ & \nodata& \nodata\\
L1689N& 8 & $0.089\pm 0.006$ & \nodata& \nodata\\
L63 & 20 & $<0.002$ & \nodata& \nodata\\
S68~FIR& 20--100 & $0.126\pm 0.055$ & 0.005& 0.010\\
S68N& 15--75 & $0.054\pm 0.010$ & \nodata& 0.009\\
\enddata
\tablenotetext{\dag}{Assumes ${\rm ^{12}C}/{\rm ^{13}C}=60$}
\tablenotetext{1}{Williams et al. 1998}
\tablenotetext{2}{Butner et al. 1995}
\tablenotetext{3}{Greason 1986}
\end{deluxetable}
\end{center}

%\clearpage
%% Table of Skewness parameters for self-absorption features
\begin{center}
\begin{deluxetable}{lcccccc}
\tabletypesize{\footnotesize}
\tablecolumns{7}
\tablewidth{0pc}
\tablecaption{Observed Asymmetries\label{tinfall}}
\tablehead{
\colhead{Source} & 
\colhead{\nh2d\tablenotemark{a}} & 
\colhead{Skewness} & 
\colhead{CS} & 
\colhead{\htwco} & 
\colhead{\hcop} & 
\colhead{\hcop} \\
\colhead{} & 
\colhead{$\jeq{1_{11}}{1_{01}}$} & 
\colhead{} & 
\colhead{\jeq{2}{1}\tablenotemark{b}} &
\colhead{$\jeq{2_{12}}{1_{11}}$\tablenotemark{b}} &
\colhead{\jeq{3}{2}\tablenotemark{c}} & 
\colhead{\jeq{4}{3}\tablenotemark{c}}}
\startdata
L1448C & Blue & $-0.014\pm 0.004$ & Blue & Red & Red & Red \\
IRAS4A & Blue & $-0.38\pm 0.09$ & Blue & Blue & Blue & Blue\\
IRAS4C & Blue & $-0.010\pm 0.004$ & \nodata & \nodata &\nodata&\nodata\\
L1551 IRS5 & None &\nodata& None & Red & None & None \\
S68N & Blue & $-0.704\pm 0.194$ &None & Blue & None & None \\
S68 FIR & Red?\tablenotemark{d} &\nodata& Blue & Red & Red & Blue \\
\enddata
\tablenotetext{a}{This paper; only the 86 GHz data displays these asymmetries}
\tablenotetext{b}{\citet{mar97}}
\tablenotetext{c}{\citet{gez97}}
\tablenotetext{d}{This is a fit by eye, not by a strict Gaussian fit
to the line features}
\end{deluxetable}
\end{center}

%%
%% The End
%%
\end{document}